\documentclass[pr,aps,praps,amsbsy]{revtex4}
\usepackage{graphics}

\begin{document}

\centerline{\bf{Volume reflection and channeling of ultrarelativistic protons in germanium bent single crystals
}}

\begin{center}
{\large\bf S.~Bellucci${}^a$ and V.A.~Maisheev${}^{b}$ }
\end{center}
\begin{center}
${}^a$ {\it
INFN-Laboratori Nazionali di Frascati,
Via E. Fermi 40, 00044 Frascati, Italy} \vspace{0.2cm}

${}^b$ {\it
Institute for High Energy Physics in National Research Center "Kurchatov Institute",
142281 Protvino, Russia} \vspace{0.2cm}
\end{center}
\vspace{2cm}

\begin{abstract}
The paper is devoted to the investigation of volume reflection and
channeling processes of  ultrarelativistic positive charged
particles moving in germanium single crystals. We demonstrate that
the choice of atomic potential on the basis of the Hartree-Fock
method and the correct choice of the Debye temperature allow us to
describe the above mentioned processes  in a good agreement with
the recent experiments. Moreover, the universal form of equations
for volume reflection presented in the paper gives a true
description of the process at a wide range of particle energies.
Standing on this study we make predictions for the mean angle
reflection  (as a function of the bending radius) of positive and
negative particles for germanium (110) and (111) crystallographic
planes.

\end{abstract}
\maketitle

In the last years there were many theoretical and experimental
investigations devoted to the study of different processes in bent
single crystals.  As an example we can point out detailed
measurements of such phenomena as volume reflection of
ultrarelativistic particles \cite{Iv,Iv1,WS}  and radiation
accompanying this process \cite{WS1}(for electron and positron
beams), focusing  and mirroring  \cite{WS2, WS3} effects and
others. As expected these investigations will be useful in using
high energy particles interactions with bent single crystals for
different applications in the accelerator practice.
  In particular, a special interest lies in the possibility to use such crystals for collimation and extraction
of proton and nuclear beams for LHC. It should be noted that most
of the investigations with bent crystals were performed with
silicon samples. However, in the article \cite{Bi} the
experimental results of the deflection of high energy protons  in
germanium single crystal were presented. In this experiment a high
enough deflection efficiency was obtained.

Recently new experiments with germanium bent single crystals were
performed. Short germanium crystals were used and such phenomena
as volume reflection and channeling of 400 GeV protons were
investigated\cite{DeS, DeS1}. However, the analysis of the results
of measurements for volume reflection is practically absent and,
for channeling, authors give calculations on the basis of Moliere
model for atomic potentials.

 Volume reflection represents a coherent scattering of relativistic charged particles in the planar electric field of the bent single crystal.
This effect was found in Monte Carlo simulations in Ref.
\cite{TV}. The analytical consideration of this process was
presented in the paper \cite{VM}.   This theory gives a very good
agreement with experimental data for (110) and (111) interplanar
fields in silicon single crystals (see Ref. \cite{WS4, BCMY}).
However, this good agreement  takes place when in calculations for
atomic electric fields the data derived from x-ray measurements
are used. Calculations based on the popular  Moliere model of the
potential give worse results.

We began   our simulations of coherent processes in germanium
single crystals from finding the mean angle of volume reflection.
The first calculations were performed for Moliere model of atomic
potentials. However, the computed mean angles of volume reflection
valuably exceed the measured ones. After this we applied the
atomic potential based on the Hartree-Fock method\cite{HF,IT} to
solve our problem.  In \cite{HF,IT} the results of calculating
atomic form factors (in the   Hartree-Fock method) are presented
in the form
\begin{equation}
F(q)= c_0 +\sum_{i=1}^4 a_i \exp(-b_i (q^2/(4\pi)),
\end{equation}
where $a_1= 16.0816,\, a_2= 6.3747,\, a_3 =3.7068,\, a_4 =3.683,\,
  b_1= 2.8509\AA^2 ,\, b_2=0.2516\AA^2, \,b_3=11.4468\AA^2,\, b_4=54.7625\AA^2,\, c_0= 2.1313$ are constants and ${\bf{q}}$ is called the scattering vector.
The relations for calculations of planar electric characteristics
can be found in Ref. \cite{BGM}. However, for calculations of the
interplanar potentials we should know the mean squared amplitude
of thermal vibrations of germanium atoms.  We can find  this value
from such an important characteristic of single crystal as its
Debye temperature.  If we know the Debye temperature $T_D$ we can
find the mean squared amplitude of atomic
 thermal vibrations\cite{TM}
\begin{equation}
\langle u^2 \rangle ={3\hbar^2 \over 4M_ak_B T_D}[1+4(T/T_D)^2\int_0^{T_D/T}{ydy\over e^y-1}]
\end{equation}
where  $M_a$ is the atom mass, $\hbar,\,k_B$ are the Planck and
Boltzmann constants and $T$ is the crystal temperature. The Debye
temperature for any substance may be determined experimentally
with the help of different methods. As pointed out in the paper
\cite{BC} there is a difference in  Debye temperatures obtained
from a specific-heat measurement and from measurements of x-ray
Bragg reflections. Besides, the authors of the paper \cite{BC}
state that the Debye temperature obtained from x-ray measurements
is coupled with the thermal atomic vibrations and hence this value
should be used in calculations of the mean square of thermal
vibrations in the crystal. For silicon and germanium the Debye
temperatures are approximately  $640^\circ$K and $360^\circ$K (and
correspond to 0.0645 and 0.068   $\AA$ of rms of thermal
vibrations) from a specific-heat measurement and $543^\circ$K and
$290^\circ$K from x-ray measurements (and correspond to 0.0747 and
0.0835  $\AA$   of rms of the thermal vibrations).

In the paper \cite{VM} it was shown that the main equations
describing the volume reflection may be presented in generalized
parameters $\Xi = \langle \alpha \rangle/\theta_c$ and $\kappa
=U_0 R/(E_0\beta^2 d)=R/R_0$ where $\langle\alpha \rangle$ is the
mean volume reflection angle, $\theta_c = \sqrt{2U_0 /( E_0
\beta^2 )}$ is the critical channeling angle (for unbent single
crystal) $R/R_0$ is the ratio of bending radius to characteristic
radius $R_0=\beta^2E_0 d/U_0$. Here $E_0$ is the particle energy,
$U_0$   is the value of  potential barrier for unbent crystal and
$d$ is the interplanar distance.

 For the first time here we  present the  explicit relation for the mean volume reflection angle in the
generalized variables

\begin{eqnarray}
 { \langle\alpha(\kappa) \rangle \over \theta_c} =  {1\over \kappa} \int_\nu^{\nu+1} d\nu \{ \int_{y_0}^{y_c} [ {1\over \sqrt{\nu/\kappa- U(y)/U_0 -y/\kappa}}
-  {1\over \sqrt{\nu/\kappa- U(y_c)/U_0 -y/\kappa}}] dy\} \, 
\end{eqnarray}
where  $\nu= E/\delta E$ and $E/U_0= \nu/\kappa=  E_0\beta^2
\theta^2_0/(2U_0)+U(y_0)/U_0+y_0/\kappa$, $\delta E= E_0\beta^2d
/R,\, y=x/d$, and the critical point $y_c$ is found from the
relation ${\nu \over \kappa} -{U(y_c)\over U_0}-{y_c\over
\kappa}=0$. The main contribution in the inner integral  brings
the area near the $y_c$-point (see \cite{VM}). This fact should be
taken into account in calculations.  For a thick single crystal ($
l_0 \gg \langle \alpha  \rangle R $, where $l_0$ is a thickness of
a crystal ) the area of large coordinates $y=x/d$ does not give
contribution in the inner integral (when $|x_0-x_c|  \gg \langle
\alpha  \rangle R^2$).

Let us call the function $U(y)/U_0$ as the normalized potential.
From Eq.(3) it follows that the value $ { \langle\alpha(\kappa)
\rangle \over \theta_c}$ depends mainly on the form of the
normalized potential (for thick enough single crystals). One can
expect that if the normalized potentials for different planes or
different single crystals differ slightly in between, then the
functions $\Xi (R/R_0)$ for corresponding cases will also differ
insignificantly in between.  We remind that $U(x)$ is a periodic
function with the period equal to $d$.

Fig. 1 illustrates the results of calculations (with the help of
Eq. (3)) of  $\Xi(R/R_0$)-functions for normalized potentials
which were shown in insert a). We select for calculations the
Hartree-Fock potential (at $T_D=290^\circ$K and at
$T_D=360^\circ$)K and  Moliere potential ((at $T_D=360^\circ$K and
at $T_D=285^\circ$K) and numbers of curves (from 1 to 4)  in the
figure take place in the corresponding order. The curve number 5
corresponds to the (110) silicon plane and it is presented for
comparison.  The calculations were done for $l_0 \approx 2$mm
\cite{DeS,DeS1}. We select $T_D=285^\circ $K (for Moliere
potential) because then $\langle u^2 \rangle ^{1/2}= 0,085\,\AA$
as in the paper \cite{DeS1}.

As we predicted, the calculated  $\Xi(R/R_0)$-functions are close
in between. Then we take the experimental data \cite{DeS1} which
was obtained for proton energy $E_0$ = 400 GeV and found
quantities $\theta_c$ and $R/R_0$ for every normalized potential.
After that we also find values $R/R_0$ for x-axis and $\Xi(R/R_0)$
for the y-axis. Thus, for every potential we  calculate the
corresponding experimental point. From the figure we see that for
Moliere potentials the obtained two points are far enough from
corresponding calculated curves (curves number 3 and 4). The curve
1 is close enough to corresponding points with number 1. We think
the fact that  both experimental points marked as 1 lie near the
curve 1 is important and means that the relative location of these
points has a natural (regular) character. These results we also
present in Table 1 for two values of the radius.

A similar procedure was made also for (111) planar potentials of
germanium single crystal. As a result we got that the experimental
data (which was measured only for one  bending radius (equal to 15
meters)) are in a good agreement with the  Hartree-Fock potential
and $T_D=290^\circ$K  ($\langle \alpha \rangle =16.33\mu$rad
(calculations)  and $\langle \alpha \rangle=  15.9 \pm0.3 \mu$rad
(experiment)).

The results of calculating the mean angle reflection presented
here for different atomic models and Debye temperatures
 and their comparison with experiment  show that a good agreement is observed  for the atomic model on  the
basis of Hartree-Fock model and Debye temperature equal to
$290^\circ$K. In principle this is an expected result. Really, in
the calculations the Hartree-Fock model was applied for every sort
of atoms, taking into account some specific atomic properties. The
recommendation for the value of $T_D$ for germanium follows from
paper \cite{BC} and stands upon x-ray measurements and theoretical
consideration.

We see that the  process of volume reflection in germanium may be
described well enough with the help of the Hartree Fock potential
at $T_D=290^\circ$K.  Besides,  our calculation of some channeling
characteristics  in this potential should also be compared with
experiments. The deflection efficiency has been measured
experimentally (see \cite{DeS1})
 as a function of the incoming angle relative to the (110) crystallographic plane of germanium.
The particles with an incoming angle in a $\pm 2\, \mu$rad interval
around a given value are selected; the efficiency of deflection
for those particles was then computed as the ratio of the
number of deflected particles to their total number.

We made the calculations of deflection efficiency for the
germanium crystal for the (110) plane and different potential
models for 2.3 and 8.2 meter bending radii. This calculation was
performed by Monte Carlo method and on the basis of a
one-trajectory approximation of a diffusion process which was
developed in Ref. \cite{BCMY}. We got the following results for
zero incoming angle: for Hartree-Fock potentials the deflection
efficiencies are 77.5 and 54 (at $T_D=290^\circ$K)   and 81 and
59.5 (at $T_D=360^\circ$K) percents, and for Moliere potential the
deflection efficiencies are 78.5 and 61  (at $T_D=285^\circ$K) and
81.5 and 65 (at $T_D=360^\circ$) percents. Here the first value in
every pair of quantities corresponds to a bending radius equal to
8.2 meter and the second one corresponds to 2.3 meter for the
bending radius. The corresponding measured values are
(74.1$\pm$1.5)\% for R=8.2 m and (56. $\pm$1.5)\% for R=2.3m.

Besides, in Table 1 the angle acceptance of  bending (110)
germanium plane is compared with the experimentally defined one.
In this case we see a satisfactory agreement.

As a result of our consideration we have demonstrated that the
choice of Hartree-Fock potential and the corresponding Debye
temperature allows one correctly to describe channeling and volume
reflection processes in germanium single crystals with agreement
with experimental data,  We present our calculations  in a
universal form which may be used at various initial conditions
(see Fig. 2). From here it follows that the mean angle of volume
reflection is $\approx 1.3$ larger for germanium  in comparison
with silicon at the same beam energy. Despite the small mean angle
of volume reflection we expect a valuable gain when using this
phenomenon in  multi strip crystal systems\cite{WS5,SB}.

In conclusion we would like to make a statement of common meaning
that the usage of more precise atomic potentials than the Moliere
one, in calculations of coherent phenomena in single crystal, is
preferable.

\newpage
\section{Figure captions}
Figure 1 The calculated universal functions $ \Xi(R/R_0) = \langle \alpha \rangle / \theta_c $ for different models of atomic potentials.
a) the planar normalized potentials, (see explanations in the text).

Figure 2 The calculated universal functions $\Xi(R/R_0)= \langle \alpha \rangle / \theta_c$ for germanium and silicon single crystals.
The upper curves are results for positively charged particles and lower ones are results for
negatively charged particles. Three points demonstrated experimental data for germanium
crystal \cite{DeS,DeS1} for R=2.3, 8.2 and 15 meters and $l_0$= 2 mm (from left to right, correspondingly).

\begin{center}
\begin{table}
\begin{tabular}{|c|c|c|c|c|c|c|c|c|c|c|c|c|c|}
\hline
$Potential$ & $U_0,eV$& $\sqrt{ \langle u^2 \rangle}, \AA $ &$ \theta_c,\,\mu rad $& $\langle \alpha_1 \rangle, \mu rad$ & $ \kappa_1  $ &$\Xi_1 $&$ \Xi_1^e$ &$\langle \alpha_2 \rangle, \mu rad   $& $\kappa_2$&$\Xi_2$&$ \Xi_2^e$ &$ \theta_1^a$&$ \theta_2^a $  \\
\hline
$M, T_D=360^\circ $ & $41.02$ & $0.068 $&$14.32$&  $ 14.74$ & $1.179$ & $ 1.029$&$0.796$&$20.02 $&$4.202$&$ 1.398$&$ 1.208$&$ 11.5$&$13.5$ \\
$M,T_D=285^\circ $ & $39.02$ & $0.085$&$13.97$&$ 13.75$  & $1.121$ &$ 0.984 $&$0.816  $&$ 19.11$& $3.997$&$1.368$&$1.238$&$ 11.1$&$13.1$ \\
$HF,T_D=360^\circ$ & $34.81$ & $0.068$ &$13.19$&$12.743$&  $1.000$ &$ 0.966 $&$0.864$&$  18.64$& $3.567$&$1.413$&$1.312$&$10.4$&$12.4$ \\
$HF,T_D=290^\circ$ & $33.04$ & $0.0835$ &$12.85$&$11.88$&  $0.949$ &$ 0.925 $&$0.887 $&$ 17.80$& $3.426$&$1.385$&$1.346$&$9.9$&$11.9$ \\
\hline
$exp.\, data$ & $-$ & $-$ &$-$&$11.4 \pm 0.2$&  $-$ &$ - $&$-$&$ 17.3 \pm 0.3$& $-$&$-$&$-$&$8.8\pm 0.8$&$ 10.9\pm0.8$ \\
\hline
\end{tabular}
\caption{ Characteristics of germanium crystal and volume
reflection and channeling processes for (110) plane. In the 1st
column the model of potential (M - Moliere, HF - Hartree-Fock) and
Debye temperature are presented. For other designations see in the
text. The lower index 1 (2) in some designations  corresponds to
the value at R=2.3 m (8.2 m). The upper index "e" in $\Xi$ values
corresponds to  experimental values. }
\end{table}
\end{center}

\begin{figure} [h]
\begin{center}
\scalebox{1.5}{\includegraphics{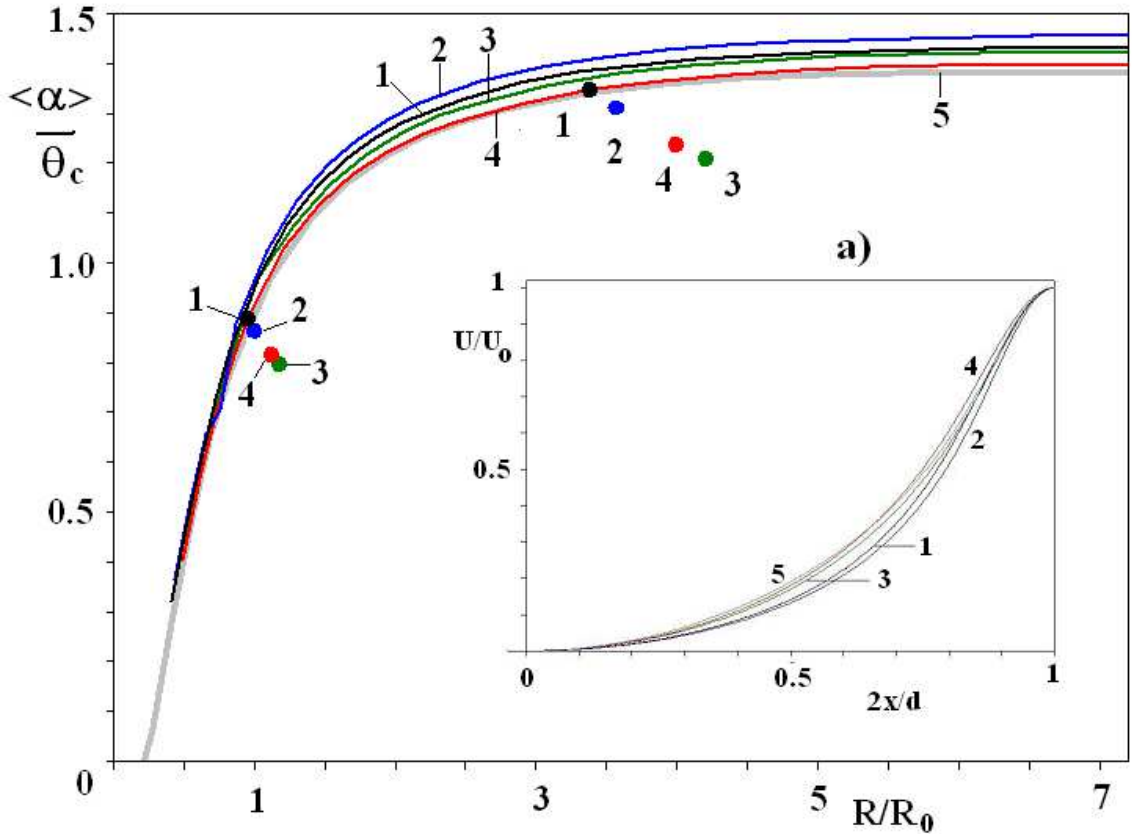}}
{\caption{
The calculated universal functions $ \Xi(R/R_0) = \langle \alpha \rangle / \theta_c $ for different models of atomic potentials.
a) the planar normalized potentials, (see explanations in the text).
    }}
\end{center}
\end{figure}
\begin{figure} [h]
\begin{center}
\scalebox{0.7}{\includegraphics{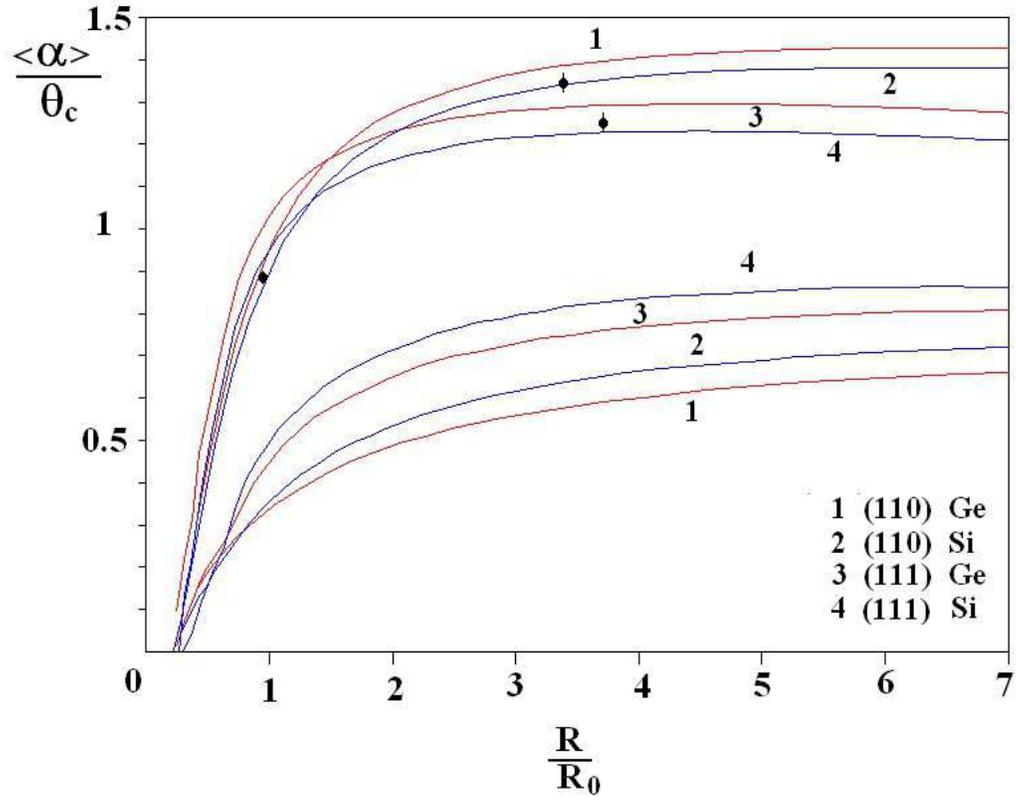}}
{\caption{
The calculated universal functions $\Xi(R/R_0)= \langle \alpha \rangle / \theta_c$ for germanium and silicon single crystals.
The upper curves are results for positively charged particles and lower ones are results for
negatively charged particles. Three points demonstrated experimental data for germanium
crystal \cite{DeS,DeS1} for R=2.3, 8.2 and 15 meters and $l_0$= 2 mm (from left to right, correspondingly).
 }}
\end{center}
\end{figure}

\end{document}